\input tables

\documentstyle[twocolumn]{article}

\newcommand{\Title}[1]{\begin{center}{\Large\bf #1}\end{center}\vskip 0.1in}
\newcommand{\Name}[1] {\begin{center}{\large    #1}\end{center}\vskip 0.1in}
\newcommand{\Abstract}[1]{\begin{center}{\large ABSTRACT}\end{center}{#1}
\vskip 0.4in}

\catcode`\@=11 
\def\Section{\@startsection {section}{1}{\z@}{-3.5ex plus -1ex minus
 -.2ex}{2.3ex plus .2ex}{\large\sc}}
\def\Subsection{\@startsection{subsection}{2}{\z@}{-3.25ex plus -1ex minus
 -.2ex}{1.5ex plus .2ex}{\large}}
\def\Subsubsection{\@startsection{subsubsection}{3}{\z@}{-3.25ex plus
-1ex minus -.2ex}{1.5ex plus .2ex}{\normalsize\it}}
\catcode`\@=12 

\def\Thebibliography#1{\section*{\large\rm REFERENCES}\list
 {[\arabic{enumi}]}{\settowidth\labelwidth{[#1]}\leftmargin\labelwidth
 \advance\leftmargin\labelsep
 \usecounter{enumi}}
 \def\newblock{\hskip .11em plus .33em minus .07em}
 \sloppy\clubpenalty4000\widowpenalty4000
 \sfcode`\.=1000\relax}

\def\Equal{\!\!\!\!=\!\!\!\!}
\def\Simeq{\!\!\!\!\simeq\!\!\!\!}

\begin{document}

\twocolumn[\begin{center}
To be published in the Proceedings of the Workshop on {\sl B Factories:
The State of the Art in Accelerators, Detectors, and Physics},
Stanford, California, April 6--10, 1992.
\end{center}]

\twocolumn[

\Title{Recent Developments in the Theory of Heavy-Quark
Decays\footnotemark}

\Name{Matthias Neubert\\
Stanford Linear Accelerator Center, Stanford, CA 94309 USA}

\Abstract{I report on recent developments in the heavy-quark effective
theory and its application to $B$ meson decays. The parameters of the
effective theory, the spin-flavor symmetry limit, and the leading
symmetry-breaking corrections to it are discussed. The results of a QCD
sum rule analysis of the universal Isgur-Wise functions that appear at
leading and subleading order in the $1/m_Q$ expansion are presented. I
illustrate the phenomenological applications of this formalism by
focusing on two specific examples: the determination of $V_{cb}$ from
the endpoint spectrum in semileptonic decays, and the study of
spin-symmetry violating effects in ratios of form factors. I also
briefly comment on nonleptonic decays.}

] 

\footnotetext{\noindent
\ \ {\small Financial support from the BASF Aktiengesellschaft and the
German National Scholarship Foundation is gratefully acknowledged.
This work was also supported by the Department of Energy, contract
DE-AC03-76SF00515.}}

\Section{Introduction}

The theoretical description of hadronic processes involving the decay
of a heavy quark has recently experienced great simplification due to
the discovery of new symmetries of QCD in the limit where
$m_Q\to\infty$ \cite{Volo,Isgu}. The properties of a hadron containing
the heavy quark then become independent of its mass and spin, and the
complexity of the hadronic dynamics results from the strong
interactions among the light degrees of freedom only. The so-called
heavy-quark effective theory (HQET) provides an elegant framework to
analyze such processes \cite{Geor}. It allows a systematic expansion of
decay amplitudes in powers of $1/m_Q$.

In the formal limit of infinite heavy-quark masses, the spin-flavor
symmetries impose restrictive constraints on weak decay amplitudes. In
the case of semileptonic transitions between two heavy pseudoscalar or
vector mesons, for instance, the large set of hadronic form factors
reduces to a single universal function, the so-called Isgur-Wise form
factor $\xi(v\cdot v')$. This function only depends on the change of
velocities that the heavy mesons undergo during the transition. It is
normalized at zero recoil ($v=v'$). This observation offers the
exciting possibility of being able to extract in a model-independent
way the weak mixing parameter $V_{cb}$ from the measurement of
semileptonic decays of beauty mesons or baryons, without limitations
arising from the ignorance of long-distance dynamics.

The heavy-quark symmetries greatly simplify the phenomenology of
semileptonic weak decays in the limit where the heavy-quark masses can
be considered very large compared to other hadronic scales in the
process. But clearly, a careful analysis of symmetry-breaking
corrections is essential for any phenomenological application. Already
at leading order in the heavy-quark expansion the spin-flavor
symmetries are violated by hard-gluon exchange. The corresponding
corrections are of perturbative nature and are known very accurately to
next-to-leading order in renormalization-group improved perturbation
theory \cite{Falk,NLO,QCD}. At order $1/m_Q$, on the other hand, one is
forced to introduce a larger set of universal form factors, which are
nonperturbative hadronic quantities such as the Isgur-Wise function
itself \cite{Luke,FGL}. These functions characterize the properties of
the light degrees of freedom in the background of the static color
source provided by the heavy quark. Their understanding is at the heart
of nonperturbative QCD. In this talk I review recent progress in this
direction. I discuss the parameters of HQET, the leading QCD and
$1/m_Q$ corrections to the infinite quark-mass limit, and some specific
applications of the effective theory to semileptonic and nonleptonic
$B$ decays.

\Section{Parameters of HQET}

The construction of HQET is based on the observation that, in the limit
$m_Q\gg\Lambda_{QCD}$, the velocity $v_\mu$ of a heavy quark is
conserved with respect to soft processes. It is then possible to remove
the mass-dependent piece of the momentum operator by a field
redefinition. To this end, one introduces a field $h_Q(v,x)$, which
annihilates a heavy quark $Q$ with velocity $v$ ($v^2=1, v_0\ge 1$), by
\cite{Geor}
\begin{equation}
   h_Q(v,x) = {(1+\rlap/v)\over 2}\,\exp(i m_Q v\!\cdot\! x)\,Q(x) .
\end{equation}
If $P_\mu$ is the total momentum of the heavy quark, the new field
carries only the residual momentum $k_\mu=P_\mu-m_Q v_\mu$, which is of
order $\Lambda_{\rm QCD}$. In the limit $m_Q\to\infty$ the effective
Lagrangian for the strong interactions of the heavy quark becomes
\begin{equation}
   {\cal{L}}_{\rm eff} = \bar h_Q\,i v\!\cdot\!D\,h_Q
   - \delta m_Q\,\bar h_Q h_Q ,
\end{equation}
where $D_\mu$ is the covariant derivative, and $\delta m_Q$ denotes the
residual mass of the heavy quark in the effective theory \cite{AMM}.

Note that there is some ambiguity associated with the construction of
HQET, since the heavy-quark mass used in the definition of the field
$h_Q$ is not uniquely defined. In fact, for HQET to be consistent it is
only necessary that $\delta m_Q$ and $k_\mu$ be of order
$\Lambda_{\rm QCD}$, {\it i.e.}, stay finite in the limit
$m_Q\to\infty$. A redefinition of $m_Q$ by a small amount $\Delta$
induces changes in these quantities. In particular, if $m_Q\to m_Q +
\Delta$, then $\delta m_Q\to\delta m_Q - \Delta$. Hence there is a
unique choice $m_Q^*$ for the heavy-quark mass in the construction of
the effective theory such that the residual mass vanishes, $\delta
m_Q=0$. This prescription provides a nonperturbative definition of the
heavy-quark mass, which has been implicitly adopted in most previous
analyses based on HQET. Yet it is important to notice that the mass
$m_Q^*$ is a nontrivial parameter of the theory. For instance, one can
associate the difference $\bar\Lambda$ between this mass and the mass
of a meson $M$ (or baryon) containing the heavy quark with the energy
carried by the light constituents. That $\bar\Lambda$ is in fact a
parameter characterizing the properties of the light degrees of freedom
becomes explicit in the relation
\begin{equation}
   \bar\Lambda = m_M - m_Q^* = {\langle\,0\,|\,\bar q\,
   (i v\!\cdot\!\overleftarrow{D})\,\Gamma\,h_Q\,| M(v)\rangle
   \over\langle\,0\,|\,\bar q\,\Gamma\,h_Q\,| M(v)\rangle} ,
\end{equation}
which can be derived from the equations of motion of HQET \cite{AMM}.
Here $\Gamma$ is an appropriate Dirac matrix such that the currents
interpolate the heavy meson $M$.

The two parameters $m_Q^*$ and $\bar\Lambda$ characterize the static
properties of the heavy quark and of the light degrees of freedom.
Their ratio determines the size of power corrections to the infinite
quark-mass limit. Assuming $\bar\Lambda\simeq 0.50$ GeV one expects
$\bar\Lambda/2 m_b^*\simeq 5\%$ and $\bar\Lambda/2 m_c^*\simeq 20\%$
for the leading power corrections relevant to processes involving $B$
or $D$ mesons, respectively. This estimate is confirmed by detailed
computations (see below).

Because of the spin-flavor symmetry the non-trivial dynamical
properties of a hadron containing the heavy quark are entirely related
to its light constituents. Consider, for instance, a transition between
two heavy mesons (pseudoscalar or vector), $M\to M'$, induced by a weak
current. At leading order in the heavy-quark expansion the associated
hadronic matrix element factorizes into a trivial kinematical part,
which depends on the mass and spin-parity quantum numbers of the
mesons, and a reduced matrix element, which describes the elastic
transition of the light degrees of freedom. Introducing spin
wave-functions by
\begin{equation}
   {\cal{M}}(v) = \sqrt{m_M}\,{(1+\rlap/v)\over 2}\,
   \cases{-\gamma_5 &; $J^P=0^-$ , \cr
          \rlap/\epsilon &; $J^P=1^-$ , \cr}
\end{equation}
one finds
\begin{equation}\label{IWF}
   \langle M'|\,\bar h_{Q'}\Gamma\,h_Q\,| M\rangle =
   - \xi(w)\,{\rm Tr}\,\big\{\,\overline{\cal{M}}'(v')\,
   \Gamma\,{\cal{M}}(v)\,\big\} ,
\end{equation}
where $w=v\cdot v'$, and $\xi(w)$ is the universal Isgur-Wise form
factor \cite{Isgu,Falk}. It measures the overlap of the wave functions
of the light degrees of freedom in the two mesons moving at velocities
$v$ and $v'$. The conservation of the vector current implies that there
is complete overlap if $v=v'$, such that at zero recoil $\xi(1)=1$.

Let us now focus on semileptonic decays of $B$ mesons. It is convenient
to define a set of heavy-meson form factors $h_i(w)$ by
\begin{eqnarray}
\lefteqn{\langle D(v') |\,V_\mu\,| \bar B(v) \rangle} \nonumber\\
   &&\!\!\!\!\!\!\!= \sqrt{m_B\,m_D}\,\Big[ h_+(w)\,(v+v')_\mu +
    h_-(w)\,(v-v')_\mu \Big] , \nonumber\\
   \nonumber\\
\lefteqn{\langle D^*(v') |\,V_\mu\,| \bar B(v) \rangle} \nonumber\\
   &&\!\!\!\!\!\!\!= i \sqrt{m_B\,m_{D^*}}\,\,h_V(w)\,
    \epsilon_{\mu\nu\alpha\beta}
    \,\epsilon^{*\nu}\,v'^\alpha\,v^\beta , \\
   \nonumber\\
\lefteqn{\langle D^*(v') |\,A_\mu\,| \bar B(v) \rangle} \nonumber\\
   &&\!\!\!\!\!\!\!= \sqrt{m_B\,m_{D^*}}\,\Big[ h_{A_1}(w)\,(w+1)\,
    \epsilon_\mu^* \nonumber\\
   &&\qquad\qquad\, -\,h_{A_2}(w)\,\epsilon^*\!\cdot\! v\,v_\mu
    - h_{A_3}(w)\,\epsilon^*\!\cdot\! v\,v'_\mu \Big] , \nonumber
\end{eqnarray}
where $V_\mu=\bar c\,\gamma_\mu\,b$ and $A_\mu=\bar c\,\gamma_\mu
\gamma_5\,b$, and $\epsilon_\mu$ is the polarization vector of the
$D^*$ meson. In the infinite quark-mass limit one finds from
(\ref{IWF})
\begin{eqnarray}\label{HQL}
   h_+(w) &\Equal& h_V(w) = h_{A_1}(w) = h_{A_3}(w) = \xi(w) ,
    \nonumber\\
   h_-(w) &\Equal& h_{A_2}(w) = 0 .
\end{eqnarray}
These relations summarize the symmetry constraints imposed on the weak
matrix elements.

The mass parameter $\bar\Lambda$ and the Isgur-Wise functions are
fundamental hadronic quantities that appear at leading order of the
heavy-quark expansion. They can only be computed using nonperturbative
techniques such as lattice gauge theory or QCD sum rules. While no
lattice results are available so far, QCD sum rules \cite{Shif} have
often been used successfully to compute hadron masses, decay constants,
and form factors. This method has been recently applied to the analysis
of form factors in HQET \cite{MNSR,SR,MNnew}. From the study of the
correlator of two heavy-light currents in the effective theory one
finds that \cite{MNSR}
\begin{equation}
   \bar\Lambda = 0.50\pm 0.07~{\rm GeV} ,
\end{equation}
corresponding to heavy-quark masses $m_b^*\simeq 4.8$ GeV and
$m_c^*\simeq 1.4$ GeV. The Isgur-Wise function is obtained from the
analysis of a three-current-correlator. The result can be parameterized
in terms of a pole-type function
\begin{equation}
   \xi(w)\simeq \bigg({2\over w+1}\bigg)^{\beta(w)} \!;~~
   \beta(w) = 2 + {0.6\over w} .
\end{equation}
It explicitly obeys the normalization condition $\xi(1)=1$ and exhibits
dipole behavior at large recoil.

\Section{Symmetry-Breaking Corrections}

{}From the fact that the mass of the charm quark is not particularly
large compared to a hadronic scale such as $\bar\Lambda$ one expects
substantial symmetry-breaking corrections to the relations (\ref{HQL}).
These have to be incorporated in any phenomenological analysis based on
HQET if the effective theory is to be more reliable than a particular
model for the hadronic form factors. The leading corrections come from
hard-gluon exchange and from terms of order $1/m_Q^*$ in the
heavy-quark expansion. I will discuss these corrections separately
below. Fortunately, it turns out that at least at zero recoil they can
be calculated in an almost model-independent way, such that reliable
predictions beyond the infinite quark-mass limit are still possible.

In order to make the heavy-quark symmetry limit and the leading
symmetry-breaking corrections to it explicit, I write
\begin{equation}\label{decomp}
   h_i(w) = \Big[ \alpha_i + \beta_i(w) + \gamma_i(w) + \ldots \Big]
   \,\xi(w) ,
\end{equation}
where $\alpha_+=\alpha_V=\alpha_{A_1}=\alpha_{A_3}=1$ and $\alpha_-=
\alpha_{A_2}=0$, the functions $\beta_i(w)$ are the short-distance
perturbative corrections, and $\gamma_i(w)$ contain the $1/m_c^*$ and
$1/m_b^*$ corrections. The ellipses represent higher-order terms.

\Subsection{QCD Corrections}

The form factors receive perturbative corrections due to the coupling
of hard gluons to the heavy quarks. The corresponding coefficients
$\beta_i(w)$ in (\ref{decomp}) are complicated functions of $w,
\alpha_s(m_c^*), \alpha_s(m_b^*)$, and the mass ratio $m_c^*/m_b^*$.
Their calculation is, however, purely perturbative and can make use of
the powerful methods of the renormalization group \cite{Falk,NLO,QCD}.
The coefficients $\beta_i(w)$ are known to next-to-leading logarithmic
order and are tabulated in Refs.~\cite{QCD}.

\Subsection{$1/m_Q^*$ Corrections}

\begin{table}
\caption{The universal form factors at leading and subleading order in
HQET.}
\smallskip
\center{
\begin{tabular}{c|c|c}
function & normalization & broken symmetries \\
\hline
$\xi(v\cdot v')$ & $\xi(1)=1$ & no \\
\hline
$\xi_3(v\cdot v')$ & no & spin, flavor \\
$\chi_1(v\cdot v')$ & $\chi_1(1)=0$ & flavor \\
$\chi_2(v\cdot v')$ & no & spin, flavor \\
$\chi_3(v\cdot v')$ & $\chi_3(1)=0$ & spin, flavor \\
\end{tabular}}
\end{table}

At subleading order in the heavy-quark expansion the currents no
longer have the simple structure as in (\ref{IWF}). Instead, there
appear higher-dimensional operators such as
\begin{equation}
   {1\over 2 m_Q^*}\,\bar h_{Q'}\Gamma\,i\,\rlap/\!D\,h_Q ,
\end{equation}
whose hadronic matrix elements give rise to new universal form factors.
In total, four additional functions are required to describe all
$1/m_Q^*$ corrections to transitions between two heavy mesons
\cite{Luke,Volker}. Their properties are collected in Table~1. The
conservation of the vector current implies that two of these functions
vanish at zero recoil. As a consequence, the hadronic form factors
$h_+(w)$ and $h_{A_1}(w)$ are protected against $1/m_Q^*$ corrections
at $w=1$. This is the content of Luke's theorem \cite{Luke}.

The subleading universal functions can again be calculated using QCD
sum rules in the effective theory. One finds \cite{MNnew}
\begin{eqnarray}\label{sumrul}
   \xi_3(w) &\Simeq& {1\over 3}\,\Big[ \xi(w) - \kappa\,(w-1) \Big] ,
    \nonumber\\
   \chi_1(w) &\Simeq& {2\over 3}\,{w-1\over w+1}\,
    \Big[ \Big( 4w + {7\over 2} \Big) \kappa - \xi(w) \Big]
    + 18\,\chi_3(w) , \nonumber\\
   \chi_2(w) &\Simeq& 0 , \nonumber\\
   \chi_3(w) &\Simeq& {\kappa\over 8}\,\Big[ 1 - \xi(w) \Big] .
\end{eqnarray}
Nonperturbative effects are contained in the Isgur-Wise function and
the parameter $\kappa\simeq 0.16$, which is proportional to the mixed
quark-gluon condensate $\langle\bar q \sigma_{\mu\nu} G^{\mu\nu}
q\rangle$. One does indeed find that the functions $\chi_1(w)$ and
$\chi_3(w)$ vanish at $w=1$. In addition, restricting to the diagrams
usually included in a sum rule analysis one finds no contribution to
the spin-symmetry violating form factor $\chi_2(w)$, and obtains the
parameter-free prediction
\begin{equation}
   \xi_3(1) = {1\over 3} .
\end{equation}
Corrections to this relation are expected to be small.

In Table~2, I show the theoretical prediction for the sum of the
symmetry-breaking corrections to the various heavy-meson form factors,
based on the most recent calculation of QCD corrections \cite{QCD} and
the above sum rule results. The relations between the corrections
$\gamma_i(w)$ and the subleading universal functions are given in
Ref.~\cite{Volker}.

\begin{table}
\caption{Total symmetry-breaking corrections $\delta_i(w)=\beta_i(w) +
\gamma_i(w)$ in \%.}
\smallskip
\center{
\begin{tabular}{c|rrrrrr}
$w$ &$\delta_+$ &$\delta_-$ &$\delta_V$ &$\delta_{A_1}$ &$\delta_{A_2}$
&$\delta_{A_3}$ \\
\hline
1.0  &2.6  &$-9.5$ &31.0 &$-1.5$ &$-34.1$ &$-1.9$ \\
1.1  &2.4  &$-9.5$ &29.6 &$-0.9$ &$-31.7$ &$-0.9$ \\
1.2  &3.1  &$-9.4$ &29.2 &0.6    &$-29.6$ &0.9    \\
1.3  &4.9  &$-9.5$ &29.8 &2.8    &$-27.6$ &3.4    \\
1.4  &7.3  &$-9.6$ &31.1 &5.7    &$-25.8$ &6.4    \\
1.5  &10.4 &$-9.7$ &33.2 &9.0    &$-24.2$ &10.0   \\
\end{tabular}}
\end{table}

\Section{Phenomenological Applications}

The theoretical results summarized in Table~2 form a solid basis for a
comprehensive analysis of semileptonic $B$ decays to subleading order
in HQET. Some specific applications, as well as the extension to
nonleptonic decays, are discussed below. I do not address here the
important issue of decay constants of heavy mesons. The reader
interested in this subject is referred to Refs.~\cite{MNSR,SR}.

\Subsection{Measurement of $V_{cb}$}

As a first application let me focus on the extraction of the
quark-mixing parameter $V_{cb}$ from the extrapolation of semileptonic
$B$ decay rates to zero recoil. This subject has been discussed in
detail in Ref.~\cite{MN}. In general, one finds that
\begin{eqnarray}
\lefteqn{\lim_{w\to 1} {1\over[w^2-1]^{1/2}}\,
    {{\rm d}\Gamma(\bar B\to D^*\ell\,\bar\nu)\over{\rm d} w}}
    \nonumber\\
   &&= {G_F^2\over 4\pi^3}\,| V_{cb} |^2\,(m_B-m_{D^*})^2\,
    m_{D^*}^3\,\eta^{*2} , \nonumber\\
   && \nonumber\\
\lefteqn{\lim_{w\to 1} {1\over[w^2-1]^{3/2}}\,
    {{\rm d}\Gamma(\bar B\to D\,\ell\,\bar\nu)\over{\rm d} w}}
    \nonumber\\
   &&= {G_F^2\over 48\pi^3}\,| V_{cb} |^2\,(m_B+m_D)^2\,
    m_D^3\,\eta^2 ,
\end{eqnarray}
with $\eta^*=\eta=1$ in the infinite quark-mass limit. Because of
Luke's theorem the decay rate for $\bar B\to D^*\ell\,\bar\nu$ is
protected against $1/m_Q^*$ corrections at zero recoil. Thus to
subleading order in HQET the coefficient $\eta^*$ deviates from unity
only due to radiative corrections. Ignoring terms of order
$1/m_Q^{*2}$, one finds that $\eta^* = 1 + \delta_{\rm QCD}^*$ with
$\delta_{\rm QCD}^*\simeq -0.01$ \cite{QCD}. On the other hand, Luke's
theorem does not apply for $\bar B\to D\,\ell\, \bar\nu$ decays because
the decay rate is helicity-suppressed at zero recoil \cite{Volker,MN}.
In this case $\eta = 1 + \delta_{\rm QCD} + \delta_{1/m_Q^*}$ with
$\delta_{\rm QCD}\simeq 0.05$ and
\begin{equation}
   \delta_{1/m_Q^*} = {\bar\Lambda\over 2} \bigg({1\over m_c^*}
    + {1\over m_b^*}\bigg) \bigg({m_B-m_D\over m_B+m_D}\bigg)^2
    \big[ 1 - 2\,\xi_3(1) \big] ,
\end{equation}
which gives $\delta_{1/m_Q^*}\simeq 0.02$. Note that the $1/m_Q^*$
corrections are suppressed by the Voloshin-Shifman factor
$[(m_B-m_D)/(m_B+m_D)]^2\simeq 0.23$ \cite{Volo}, and that the
corrections to the sum rule prediction $\xi_3(1)=1/3$ are expected to
be small. Since the canonical size of $1/m_Q^{*2}$ corrections is
$1-5$\%, I thus conclude that the theoretical uncertainty in $\eta$ is
comparable to that in $\eta^*$. Hence one should extract $V_{cb}$ from
both decay modes, using the theoretical numbers
\begin{equation}
   \eta^*\simeq 0.99 ,~~ \eta\simeq 1.07 ,
\end{equation}
which are expected to have an accuracy of better than 5\%.

Until now such an analysis has only been performed for
$\bar B\to D^*\ell\,\bar\nu$ \cite{MN}. Using the updated value for
the total branching ratio as measured by CLEO, ${\rm B}(\bar B\to D^*
\ell\,\bar\nu)=4.4\pm 0.5$\% \cite{Sheldon}, I find
\begin{equation}
   V_{cb} = 0.040\pm 0.005
\end{equation}
for $\tau_B=1.3$ ps.

\Subsection{Ratios of Form Factors}

It has been emphasized in Ref.~\cite{MNnew} that a measurement of
spin-symmetry-breaking effects in ratios of the various form factors
that describe $\bar B\to D^*\ell\,\bar\nu$ transitions would not only
offer the possibility of a nontrivial test of HQET beyond the leading
order, but also provide valuable information about nonperturbative QCD.
In the limit where the lepton mass is neglected, two axial form
factors, $A_1(q^2)$ and $A_2(q^2)$, and one vector form factor,
$V(q^2)$, are observable in these decays. The ratios
\begin{eqnarray}
   R_1 &\Equal& \bigg[ 1 - \displaystyle{q^2\over(m_B+m_{D^*})^2}
    \bigg]\,{V(q^2)\over A_1(q^2)} , \nonumber\\
   \\
   R_2 &\Equal& \bigg[ 1 - \displaystyle{q^2\over(m_B+m_{D^*})^2}
    \bigg]\,{A_2(q^2)\over A_1(q^2)} \nonumber
\end{eqnarray}
become equal to unity in the infinite quark-mass limit and are thus
sensitive measures of symmetry-breaking effects.

To subleading order in HQET, I write
\begin{equation}
   R_i = 1 + \varepsilon_i^{\rm QCD}
   + \varepsilon_i^{1/m_Q^*} ;~~ i=1,2.
\end{equation}
The theoretical prediction for $\varepsilon_i$ as a function of $q^2$
is shown in Table~3. I propose a measurement of these quantities as an
ideal test of the heavy-quark expansion for $b\to c$ transitions. In
particular, note that the large values of $R_1$ result from both large
QCD and $1/m_Q^*$ corrections. The latter ones are to a large extent
model-independent since the subleading universal functions only appear
in the $1/m_b^*$ terms \cite{MNnew}. Thus the sizeable deviation of
$R_1$ from the symmetry limit $R_1=1$ is an unambiguous prediction of
HQET. A measurement of this ratio with an accuracy of 10\% could
provide valuable information about the size of higher-order
corrections. The ratio $R_2$, on the other hand, receives only very
small QCD corrections and is sensitive to the subleading form factors
$\xi_3(w)$ and $\chi_2(w)$. It can be used to test the sum rule
predictions (\ref{sumrul}). For the practical feasibility of such tests
it seems welcome that the theoretical predictions for both ratios are
almost independent of $q^2$ ($R_1\simeq 1.3$ and $R_2\simeq 0.9$), such
that it suffices to measure the integrated ratios.

\begin{table}
\caption{Theoretical predictions for the symmetry-breaking corrections
$\varepsilon_i$ in \%.}
\smallskip
\center{
\begin{tabular}{c|cc|cr}
$q^2~[{\rm GeV^2}]$ &$\varepsilon_1^{\rm QCD}$
&$\varepsilon_1^{1/m_Q^*}$ &$\varepsilon_2^{\rm QCD}$
&$\varepsilon_2^{1/m_Q^*}$ \\
\hline
10.69 &12.0 &19.1 &0.5 &$-11.0$ \\
8.57  &11.7 &18.2 &0.5 &$-10.3$ \\
6.45  &11.3 &17.5 &0.5 &$-9.6$  \\
4.33  &11.0 &16.8 &0.5 &$-8.9$  \\
2.21  &10.7 &16.2 &0.5 &$-8.3$  \\
0.09  &10.4 &15.6 &0.5 &$-7.7$  \\
\end{tabular}}
\end{table}

\Subsection{Nonleptonic Decays}

As a final application, let me briefly comment on nonleptonic two-body
decays of $B$ mesons. In this case, the heavy-quark symmetries do not
yield relations as restrictive as those for semileptonic transitions.
One still has to rely on the factorization hypothesis, under which the
complicated hadronic matrix elements of the weak Hamiltonian reduce to
products of decay constants and matrix elements of current operators,
which are of the same type as those encountered in semileptonic
processes. It is at this stage that the heavy-quark symmetries can be
advantageously incorporated, leading to essentially model-independent
predictions for the factorized decay amplitudes. This provides for the
first time a clean framework in which to test factorization. The
procedure would be as follows: One extracts the Isgur-Wise function
from data on semileptonic $B$ decays and incorporates the leading
symmetry-breaking corrections as discussed above. This determines the
functions $h_i(w)$, which suffice to compute all matrix elements that
appear in the factorized decay amplitudes for nonleptonic processes.
Besides decay constants, these amplitudes contain two parameters, $a_1$
and $a_2$, which are related to the Wilson coefficients of the
effective Hamiltonian. They would be universal numbers if factorization
were exact. In cases where the relevant decay constants are known, a
case-by-case determination of $a_1$ or $a_2$ provides a test of
factorization. In other cases, one may rely on factorization to obtain
estimates for yet unknown decay constants such as $f_{D_S}$. Both
strategies have been pursued by various authors, and we refer the
interested reader to the literature \cite{nonlep}.

\Section{Conclusions}

I have presented a short overview of recent developments in the theory
of heavy-quark decays. The spin-flavor symmetries that QCD reveals for
heavy quarks lead to relations among the hadronic form factors which
describe semileptonic decays of heavy mesons or baryons. The
heavy-quark effective theory provides a convenient framework for the
analysis of such processes. It allows a separation of short- and
long-distance phenomena in such a way that the nontrivial dynamical
information is parameterized in terms of universal functions, which
describe the properties of the light degrees of freedom in the
background of the static color source provided by the heavy quark.
These functions are fundamental, nonperturbative quantities of QCD. I
have presented explicit expressions for them obtained from QCD sum
rules. In the near future, similar results should be obtainable from
lattice gauge theory.

If the leading symmetry-breaking corrections are taken into account,
the heavy-quark effective theory forms a solid, almost
model-independent basis for an analysis of many weak decay processes. I
have discussed the determination of $V_{cb}$ from the endpoint spectrum
in semileptonic $B$ decays, and the study of symmetry-breaking effects
in ratios of form factors, which offers nontrivial tests of the
heavy-quark expansion beyond leading order. I have also emphasized that
the use of the spin-flavor symmetry may provide a cleaner basis for
tests of factorization in nonleptonic two-body decays of $B$ mesons.

\bigskip
{\it Acknowledgement:} Part of the work reported here has been done in
a most enjoyable collaboration with A. Falk and M. Luke.

\begin{Thebibliography}{99}

{\small

\bibitem {Volo}
M.B. Voloshin and M.A. Shifman, Yad.\ Fiz.\ {\bf 45} (1987) 463
[Sov.\ J.\ Nucl.\ Phys.\ {\bf 45} (1987) 292], Yad.\ Fiz.\ {\bf 47}
(1988) 511 [Sov.\ J.\ Nucl.\ Phys.\ {\bf 47} (1988) 511].

\bibitem {Isgu}
N. Isgur and M.B. Wise, Phys.\ Lett.\ {\bf B232} (1989) 113,
Phys.\ Lett.\ {\bf B237} (1990) 527.

\bibitem {Geor}
H. Georgi, Phys.\ Lett.\ {\bf B240} (1990) 447.

\bibitem {Falk}
A.F. Falk, H. Georgi, B. Grinstein and M.B. Wise, Nucl.\ Phys.\ {\bf
B343} (1990) 1.

\bibitem {NLO}
G.P. Korchemsky and A.V. Radyushkin, Nucl.\ Phys.\ {\bf B283} (1987)
342; G.P. Korchemsky, Mod.\ Phys.\ Lett.\ {\bf A4} (1989) 1257; X. Ji
and M.J. Musolf, Phys.\ Lett.\ {\bf B257} (1991) 409; D.L. Broadhurst
and A.G. Grozin, Phys.\ Lett.\ {\bf B267} (1991) 105.

\bibitem {QCD}
M. Neubert, Nucl.\ Phys.\ {\bf B371} (1992) 149, Heidelberg preprint
HD--THEP--91--30 (1991), to be published in Phys.\ Rev.\ {\bf D}.

\bibitem {Luke}
M.E. Luke, Phys.\ Lett.\ {\bf B252} (1990) 447.

\bibitem {FGL}
A.F. Falk, B. Grinstein and M.E. Luke, Nucl.\ Phys.\ {\bf B357} (1991)
185.

\bibitem {AMM}
A.F. Falk, M. Neubert and M.E. Luke, SLAC preprint SLAC--PUB--5771
(1992).

\bibitem {Shif}
M.A. Shifman, A.I. Vainshtein and V.I. Zakharov, Nucl.\ Phys.\
{\bf B147} (1979) 385, Nucl.\ Phys.\ {\bf B147} (1979) 448.

\bibitem {MNSR}
M. Neubert, Phys.\ Rev.\ {\bf D45} (1992) 2451, SLAC preprint
SLAC--PUB--5770 (1992), to be published in Phys.\ Rev.\ {\bf D46}.

\bibitem {SR}
A.V. Radyushkin, Phys.\ Lett.\ {\bf B271} (1991) 218; E. Bagan,
P. Ball, V.M. Braun and H.G. Dosch, Phys.\ Lett.\ {\bf B278} (1992) 457.

\bibitem {MNnew}
M. Neubert, SLAC preprint SLAC--PUB--5826 (1992).

\bibitem {Volker}~
M. Neubert and V. Rieckert, Heidelberg preprint HD--THEP--91--6 (1991),
to be published in Nucl.\ Phys.\ {\bf B}.

\bibitem {MN}~
M. Neubert, Phys.\ Lett.\ {\bf B264} (1991) 455.

\bibitem {Sheldon}~
S. Stone, in {\sl B Decays}, edited by S. Stone (World Scientific,
Singapore, 1992).

\bibitem {nonlep}
D. Bortoletto and S. Stone, Phys.\ Rev.\ Lett.\ {\bf 65} (1990) 2951;
J.L. Rosner, Phys.\ Rev.\ {\bf D42} (1990) 3732;
M. Neubert, V. Rieckert, B. Stech and Q.P. Xu, in {\sl Heavy Flavours},
edited by A.J. Buras and M. Lindner, Advanced Series on Directions in
High Energy Physics (World Scientific, Singapore, in press);
M. Neubert, in {\sl Proceedings of the Joint International
Lepton-Photon Symposium and Europhysics Conference on High Energy
Physics}, Geneva, Switzerland, 1991, edited by S. Hegarty, K. Potler,
and E. Quercigh (World Scientific, Singapore, 1992); H. Albrecht {\it et
al.}, Z.\ Phys.\ {\bf C54} (1992) 1.

}

\end{Thebibliography}

\end{document}